\documentclass[twocolumn,showpacs,preprintnumbers,amsmath,amssymb,showkeys]{revtex4}
\usepackage{graphicx}% Include figure files
\usepackage{dcolumn}% Align table columns on decimal point
\usepackage{bm}% bold math

\newcommand{\eps}{\varepsilon}
\newcommand{\ph}{\varphi}

\newcommand{\trn}{^{\rm\scriptscriptstyle T}}

\newcommand{\mR}{\mathbb R}
\newcommand{\mZ}{\mathbb Z}
\newcommand{\inr}{\!\in \mR}

\DeclareMathOperator{\sign}{sign} 
\DeclareMathOperator{\rank}{rank}
\DeclareMathOperator{\dist}{dist}
\newcommand{\be}{\begin{equation}}
\newcommand{\ee}{\end{equation}}

\renewcommand{\dfrac}{\frac}
\begin{document}

\preprint{APS}

\title{ Chaotic Observer-based Synchronization  \\
Under Information Constraints}% Force line breaks with \\

\author{Alexander L. Fradkov, Boris Andrievsky}
  \email{{alf,bandri}@control.ipme.ru}%
\affiliation{Institute for Problems of Mechanical Engineering,
Russian Academy of Sciences, \\ 61, Bolshoy V.O. Av., 199178,
Saint Petersburg, Russia %
}%
\author{Robin J. Evans}
\email{r.evans@ee.unimelb.edu.au.}%
\affiliation{National ICT Australia,\\%
Department of Electrical and electronic Engineering,\\%
University of Melbourne, Victoria, 3010, Australia%
}%
\date{\today}% It is always \today, today,

\begin{abstract}
Limit possibilities of observer-based synchronization systems
under information constraints (limited information capacity of the
coupling channel) are evaluated.
 We give theoretical analysis for
multi-dimensional drive-response systems represented in the  Lurie
form (linear part plus nonlinearity depending only on measurable
outputs). It is shown that the upper bound of the limit
synchronization error (LSE) is proportional to the upper bound of
the transmission error. As a consequence, the upper and lower
bounds of LSE are proportional to the maximum rate of the coupling
signal and inversely proportional to the information transmission
rate (channel capacity). Optimality of the binary coding for
coders with one-step memory is established. The results are
applied to synchronization of two chaotic Chua systems coupled via
a channel with limited capacity.
\end{abstract}

%Use showkeys class option if keyword

\maketitle

\section{Introduction}
Chaotic synchronization has attracted the attention of researchers since
the 1980s \cite{Romanovsky83,Fujisaka83,Afraimovich86} and is
still an area of active research \cite{Pikovsky01,Kurths02,Freitas05}.
Recently information-theoretic concepts  were applied to analyze
and quantify synchronization \cite{Stojanovski97,Palus01,Shabunin02,Corron03,%
Kurths05}. In  \cite{Palus01,Shabunin02} mutual
information measures were introduced for evaluating the degree of
chaotic synchronization. In \cite{Stojanovski97,Corron03} the methods of symbolic dynamics were used to relate synchronization precision to
capacity of the information channel and to the entropy of the drive
system. Baptista and Kurths \cite{Kurths05} introduced the concept
of a chaotic channel as a medium formed by a network of chaotic
systems that enables information from a source to pass from one system (transmitter) to another system (receiver). They
characterized a chaotic channel by the mutual information
(difference between the sum of the positive Lyapunov exponents
corresponding to the synchronization manifold and the sum of
positive exponents corresponding to the transverse manifold).
However, in existing papers  limit possibilities for the precision
 of controlled synchronization have not been analyzed.

Recently the limitations of control under constraints imposed by
a finite capacity information channel have been investigated in
detail in the control theoretic literature, see \cite{WongBrockett_AC97e,NairEvans_Aut03,NairEvans_SIAM04,Nair04,BazziMitter_InTh05} and references
therein. It was shown that stabilization under information constraints is possible if and only if the capacity of the information channel exceeds the entropy production of the system at the equilibrium \cite{NairEvans_Aut03,NairEvans_SIAM04,Nair04}. In
\cite{Lloyd00,Lloyd04} a general statement was proposed, claiming
that the difference between the entropies of the open loop and the
closed loop systems cannot exceed the information introduced by
the controller, including the transmission rate of the information
channel. However, results of the previous works on control
system analysis under information constraints do not apply to 
synchronization systems since in a synchronization problem
trajectories in the phase space  converge to a set (a manifold)
rather than to a point, i.e. the problem cannot be reduced to simple
stabilization.

In this paper we establish limit possibilities of observer-based
synchronization systems under information constraints.
Observer-based synchronization systems are used when only one
phase variable is available for measurement and coupling. Such systems are
well studied without information constraints
\cite{Morgul96,NijmeijerMareels97,Nijmeijer01}. Here we present a 
theoretical analysis for $n$-dimensional drive-response systems
represented in the so called Lurie form (linear part plus
nonlinearity, depending only on measurable
outputs). It is shown that the upper bound of the limit
synchronization error (LSE) is proportional to the upper bound of
the transmission error. As a consequence, the upper and lower
bounds of LSE are proportional to the maximum rate of the coupling
signal and inversely proportional to the information transmission
rate (channel capacity). Optimality of the binary coding for
coders with one-step memory is established. 

 Note also that it was claimed in some papers that if the capacity of the
channel is larger than the Kolmogorov-–Sinai entropy of the driving
system, then the synchronization error can be made arbitrarily
small. Such a claim is based upon the noisy channel theorem of
Shannon information theory stating that, if the source entropy is
smaller than the channel capacity, then the data generated by the
source can be transmitted over the channel with negligible
probability of error. However, it is known that to transmit data
with a sufficiently small error a sufficiently  long codeword and
a long transmission time is needed. During such a long time an
unstable chaotic trajectory may go far from the its predicted
value and synchronization may fail. Therefore analysis of the
system precision under information constraints requires more
subtle arguments which are provided in this paper based on
Lyapunov functions and coding analysis.

\section{Description of observed-based synchronization system}

To simplify exposition we will consider unidirectionally coupled
systems in the so-called Lurie form: right-hand sides are split into a linear part and a nonlinearity
vector depending only on the measured output. Then the drive system is
modelled  as follows: 
\begin{align} 
\dot x=Ax+\varphi (y),~y=Cx, \label{1}
\end{align} 
where $x$ is an $n$-dimensional (column) vector of state
variables, $y$ is the scalar output (coupling) variable, $A$ is an
$n\times n$-matrix, $C$ is $n\times 1$ (row) matrix, $\varphi(y)$
is a continuous nonlinearity. We assume that the vector of initial
conditions $x_0=x(0)$ belongs to a bounded set $\Omega$ such that
all the trajectories of the system (\ref{1}) starting in $\Omega$
are bounded. Such an assumption is typical for chaotic systems.

The response system is described as a nonlinear observer 
\begin{align}
\dot{\hat x}=A\hat x+\varphi (y)+L(y-\hat y),~\hat y=C\hat x,
\label{2} \end{align} 
where $L$ is the vector of the observer parameters
(gain). Apparently, the dynamics of the state error vector
$e(t)=x(t)-\hat x(t)$ is described by a linear equation 
\begin{align} 
\dot e=A_L e,~y=Cx, \label{3}\end{align} 
where $A_L=A-LC$. 
%Moreover, the dynamics of drive system should be taken into account too. 

As is known from control theory, e.g. \cite{Glad00}, if the pair
$(A,C)$ is observable, i.e. if $\rank(C\trn,A\trn C\trn,\dots,
(A\trn)^{n-1} C\trn) =n$, then there exists $L$ providing the
matrix $A_L$ with any given eigenvalues. Particularly, all 
eigenvalues of $A_L$ can have negative real parts, i.e. the system
(\ref{3}) can be made asymptotically stable and $e(t)\to 0$ as
$t\to \infty$. Therefore, in the absence of measurement and
transmission errors the synchronization error decays to zero.

Now let us take into account transmission errors.
We assume that the observation signal $y(t)$ is coded with symbols from a finite alphabet at discrete sampling time instants $t_k=k T_s, k=0,1,2,\ldots$, where $T_s$ is the sampling time. Let the coded symbol $\bar y_k=\bar y(t_k)$ be
transmitted over a digital communication channel with a finite capacity. To simplify the analysis,
we assume that the observations are not corrupted by observation noise; transmissions delay 
and transmission channel distortions may be neglected  and the coded  symbols are available at the receiver side at the same sampling instant $t_k=kT_s$.

Assume that {\it zero-order extrapolation} is used to convert the digital sequence $\bar {y}_k$ to the continuous-time input of the response  system $\bar y(t)$, namely, that $\bar y(t)=\bar{y}_k$ as $ kT_s \le t< (k+1)T_s $. Then the {\it transmission error} is defined as follows: 
\begin{align}
\delta_y(t)= y(t)- \bar y(t).
\label{dy}
\end{align}

In presence of the transmission error, equation \eqref{3} takes the form
\begin{align}
\dot e=Ae+\ph(y)-\ph\big(y+\delta_y(t)\big)-L_y\delta_y(t)
\label{10}
\end{align}

Our goal is to evaluate limitations imposed on the synchronization precision by limited transmission rate. To this end introduce an upper bound of the limit synchronization error $Q=\sup \overline{\lim\limits_{t\to\infty}} \|e(t)\|$, where $e(t)$ is from \eqref{10}, $\|\cdot\|$ denotes the Euclidian norm of a vector, and the supremum is taken over all admissible transmission errors.
In the next two sections we describe encoding and decoding procedures and evaluate the set of admissible transmission errors $\delta_y(t)$ for the optimal choice of coder parameters. It will be shown that $\delta_y(t)$ is bounded and does not tend to zero.

  \section{Coding procedures }

At first, consider the memoryless (static) encoder with uniform discretization and constant range. 
For a given real number $M>0$ and positive integer $\nu\in\mZ$ define a uniform scaled coder to be a discretized map $q_{\nu,M}:\mR\to\mR$  as follows.  
Introduce the {\it range   interval} ${\cal I}=[-M, M]$ of length $2M$ and the {\it discretization interval} of length $\delta=2^{1-\nu}M$ and define the coder function $q_{\nu,M}(y)$ as
\begin{align}
\label{qu1}
q_{\nu,M}(y)=
\begin{cases}
\delta\cdot\langle\delta^{-1}y\rangle,&\text{if}~~|y|\le M,\\
M\sign(y), &\text{otherwise},
\end{cases}
\end{align}
where $\langle\cdot\rangle$ denotes round-up to the nearest integer function,  $\sign(\cdot)$ is the signum function: $\sign(y)=1$, if $y\ge 0$, $\sign (y)=-1$, if $y<0$. Evidently, $|y-q_{\nu,M}(y)|\le \delta/2$ for all $y$ such that $y: |y|\le M+\delta/2$ and all values of $q_{\nu,M}(y)$ belong to the range interval ${\cal I}$. Notice that 
the interval ${\cal I}$ is equally split into $2^\nu$ parts. Therefore,
the cardinality of the mapping $q_{\nu,M}$ image is equal to $2^\nu+1$, and each codeword symbol contains $R=\log_2(2^\nu+1)$ bits of 
information. Thus, the discretized 
output of the considered encoder is found as $\bar y=q_{\nu,M}(y)$. 
We assume that the encoder and decoder  make decisions based on the same information. 

In a number of papers more sophisticated encoding schemes have been proposed and analyzed, see \cite{NairEvans_Aut03,BrockettLiberzon_AC00,%
Liberzon03,TatikondaMitter_AC04} for example. The underlying idea for coders of this kind is to reduce the range parameter $M$, replacing the symmetric range interval ${\cal I}$ by the interval ${\cal Y}_{k+1}$, covering some area around the predicted value for the $(k+1)$th observation $y_{k+1}$, $ y_{k+1}\in {\cal Y}_{k+1}$.  If the length of ${\cal Y}_{k+1}$ is small compared with the full range of possible measured output values $y$, then there is an opportunity to  reduce the range parameter $M$ and, consequently, to  decrease the coding interval $\delta$ preserving the bit-rate of transmission.  To realize this scheme, memory should be introduced into the encoder. Using such a ``zooming'' strategy it is possible to increase coder accuracy in the steady-state mode, and, at the same time, to prevent coder saturation at the beginning of the process. 

%This means that the quantizer range $M$ is updated during the time and a time-%varying  quantizer (with different values of $M$ for each instant, $M=M_k$) is %used. 
%The values of $M_k$ may be precomputed (the {\it time-based} zooming), or, %alternatively, current quantized measurements may be used at each step to %update $M_k$  (the {\it event-based} zooming). 

In this paper we use a simple version of such an encoder having one-step memory and time-based zooming. To describe it we introduce the sequence of {\it central numbers} $c_k$, $k=0,1,2,\dots$ with initial condition $c_0=0$.
At step $k$  the encoder compares the current measured output $y_k$ with the number $c_k$, forming the deviation signal $\partial y_k=y_k-c_k$. Then this signal is discretized with a given $\nu$ and 
$M=M_k$ according to \eqref{qu1}. The output signal 
\begin{align}
\label{dybar}
\bar{\partial} y_k=
q_{\nu,M_k}(\partial y_k)
\end{align}
 is represented as an $R$-bit information symbol from the coding alphabet
and transmitted over the communication 
channel to the decoder. Then the central number $c_{k+1}$ and the range parameter $M_k$ are renewed based on the available information about the driving system dynamics. 
%Assuming that the driving system output $y$ changes at a slow rate,
% i.e. $y_{k+1}\approx y_{k}$. 
We use the following update  algorithms: 
\begin{align}
\label{ck}
c_{k+1}=c_k+\bar{\partial} y_k, \quad c_0=0,~~k=0,1,\dots ,
\end{align}
\begin{align}
\label{mk}
M_{k}=(M_0- M_{\infty})\rho^k+ M_{\infty}, ~~k=0,1,\dots ,
\end{align}
where $0<\rho\le 1$ is the decay parameter, $M_{\infty}$ stands for the limit value of $M_k$. The initial value $M_0$ should be large enough to capture all the region of possible initial values of $y_0$. 

The equations \eqref{qu1}, \eqref{dybar}, \eqref{mk} describe the encoder algorithm. The same algorithm is realized by the decoder.
Namely, the decoder calculates the variables $\tilde{c}_k$, $\tilde{M}_k$ based on received codeword flow similarly to $c_k$, $M_k$.

\section{Coder optimization}

We now find a relation between the transmission rate and the achievable accuracy of the coder--decoder pair, assuming that the growth rate of $y(t)$ is uniformly bounded. Obviously, the exact bound $L_y$ for the rate of $y(t)$ is $L_y=\sup\limits_{x\in\Omega}{|C\dot x|}$, where $\dot x$ is from \eqref{1}. 
To analyze the coder--encoder accuracy, evaluate the upper bound $\Delta=\sup\limits_{t}|\delta y(t)|$ of the transmission error $\delta_y(t)=y(t)-\bar y(t)$. Consider the sampling interval $[t_k,t_{k+1}]$. It is clear that $|\delta_y(t_k)|$ does not exceed $\delta/2$. Additionally, the error may increase from $t_k$ to $t_{k+1}$ due to change of $y(t)$ by a value not exceeding $\sup\limits_{t_k<t<t_{k+1}}|y(t)-y(t_k)|$ $\le \int\limits_{t_k}^{t_{k+1}}{|\dot{y}(\tau)|\,d\tau}$ $ \le\int\limits_{t_k}^{t_{k+1}}{L_y\,d\tau}$ $=L_yT_s$. Therefore the total transmission error for each interval $[t_k,t_{k+1}]$ satisfies the inequality:
\begin{align}
|\delta_y(t)|\le \delta/2+L_yT_s
\label{delles}
\end{align}
Inequality \eqref{delles} shows that in order to meet the inequality $|\delta_y(t)|\le \Delta$ for all $t$, the sampling interval $T_s$ should satisfy condition
\begin{align}
T_s<\Delta/L_y.
\label{ts}
\end{align} 

Furthermore, if the condition \eqref{ts} holds, the given bound for the coding error will be guaranteed if the coding interval $\delta$ is appropriately chosen, namely, $\delta<2\Delta-2L_yT_s$. It provides the lower bound for the transmission bit-per-step rate $R=\log_2(2M/\delta+1)$: its value should not be less than $\log_2\left(\dfrac{M}{\Delta-L_yT_s}+1\right)$. Therefore, the coder with range $2M$, coding interval $\delta$ and sampling period $T_s$ ensures the total transmission error $\Delta$ if \eqref{ts} holds
and the transmission rate satisfies inequality
\begin{align}
\label{2!}
R\ge\log_2\left(\dfrac{M}{\Delta-L_yT_s}+1\right).
\end{align}
It follows from \eqref{ts}, \eqref{2!} that if $T_s$ is sufficiently small and $R$ is sufficiently large, then an arbitrarily small value of $\Delta$ can be assured.

Let us now optimize the coder parameters to achieve the minimum bound for the error $\Delta$. As seen from \eqref{2!}, reduction in the coder range $2M$ results in a reduction in the transmission rate $R$ and channel capacity $R^*$. On the other hand, to prevent coder saturation, $M$ should not be less than $\sup\limits_{k\in \mZ}|\delta_y(t_k)|-\delta/2=\Delta-\delta/2$. Taking into account that $\delta=2^{1-\nu}M$, we arrive at the following formula for the minimal admissible range:
\begin{align}
M=\dfrac{2^\nu}{2^\nu+1}\Delta.
\label{3!}
\end{align}

{\it Remark 1.} At the initial stage of the system evolution the error $|\delta_y|$ may exceed the bound $\Delta$, because the initial value $y(0)$ is not known. This leads to the transient mode of the system behavior. The zooming strategy may be efficient at this stage, providing the following recip\'{e} for the choice of the coder parameters: $M_0\!=\! M_y$, $M_\infty={2^\nu}\Delta /({2^\nu+1})$, where $M_y=\sup\limits_{\scriptstyle x_0\in \Omega }|y(t)|$.

Now optimize the coder w.r.t. $T_s$.
Consider the steady-state mode when $|y(t)-c_k|\le \Delta$ for each time interval $t\in[t_k, t_{k+1})$. Let $M$ be found from \eqref{3!}. Introduce a real number $\eps$ as $\eps=L_yT_s/\Delta$, (evidently, $0<\eps<1$) and rewrite the lower bound $R^*$ for $R$ in the form
\begin{align}R^*&=\log_2\bigg(\dfrac{2^\nu}{(2^\nu+1)(1-\eps)}+1\bigg).
\label{roteps}\end{align}
Defining the {\it bit-per-second} rate $\bar R={R}/{T_s}$ and its lower bound $\bar R^*$ we have from \eqref{roteps}:
\begin{align}\bar R^*&=\dfrac{L_y}{\eps\Delta}\log_2\left(\dfrac{2^\nu}{(2^\nu +1)(1-\eps)}+1\right).
\label{rps}\end{align}

Now the optimization of the coder is reduced to the following minimization problem: Find $(\eps^*,\nu^*)=\arg\min\limits_{ {\scriptstyle \eps\in(0,1)} \atop {\scriptstyle \nu\in\mZ}}{\bar R(\eps,\nu)}$.
Since the right-hand side of \eqref{rps} is strictly growing in $\nu$, the optimal value of $\nu$ is $\nu^*=0$. This means that
the {\it binary} coding scheme gives the optimal transmission rate $R^*=1$ bit per step, which yields $M^*=\Delta/2$ as an optimal value for $M$, and the {\it signum} function as an optimal coder function: $\bar y=\dfrac{\Delta}{2}\sign y.$

For the optimal value of $\nu$ we have $\bar R^*=(L_y/\Delta) r(\eps)$, where
\begin{align}r(\eps)&=\dfrac{1}{\eps}\log_2\left(\dfrac{1}{2(1-\eps)}+1\right).
\label{rot1}\end{align}

Let $r^*=\min\limits_{0<\eps<1}r(\eps)$. It is easy to see that this mimimum exists and satisfies the transcendental equation $dr(\eps)/d\eps=0$. Numerical one-dimensional minimization
yields $r^*=r(\eps^*)\approx 1.688$, where $\eps^*\approx 0.5923$. 

Therefore, the
optimal sampling time $T_s^*$ is
\begin{align}
T_s^*= \eps^*\dfrac{\Delta}{L_y}.
\label{topt}
\end{align}
Then the minimal channel bit-rate $\bar{R}^*=1/T_s$ is
\begin{align}
\bar{R}^*= r^*\dfrac{L_y}{\Delta},
\label{ropt}
\end{align}
and this bound is tight for the considered class of coders.
The relation \eqref{ropt} can be rewritten as 
\begin{align}
\bar{R}\Delta \ge r^*L_y,
\label{ropt2}
\end{align}
playing the role of an uncertainty relation between the propagation rate of information and the transmission error.

{\it Remark 2.} The limits of synchronization error may be different if a more sophisticated coder is used, e.g. a first-order coder with linear extrapolation of the signal or $n$th order coder with predictive model of the drive system. For example, if a full order observer is admitted at the transmitter side and there are $n$ channels for simultaneous transmission of the $n$-dimensional vector $\hat{x}(t_k)$ of  estimates of the drive system state, then the coder can calculate the best estimate $\hat{x}(t_{k+1})$ and choose $c_{k+1}=C\hat{x}(t_{k+1})$. In this case the prediction error for a binary coder will be determined by the divergence rate of neighboring trajectories, i.e. relation \eqref{delles} should be replaced by $ \dfrac{\Delta}{2}\exp(hT_s)<\Delta$, where $h>0$ is the upper Lyapunov exponent of the chaotic drive system. This yields the bound $T_s<\ln(2\|C\|)/h$, instead of the bound \eqref{ts}. For the transmission rate it gives the necessary condition $R^*> h/\ln(2\|C\|)$ instead of the lower bound $R^*>L_y/\Delta$ following from \eqref{ts}. If the condition $R^*>h/\ln(2\|C\|)$ holds, then the upper bound for transmission error $\Delta$ will decrease at each sampling interval $[t_k,t_{k+1})$ in $h/\big(R^*\ln(2\|C\|)\big)$ times and, therefore, will converge to zero exponentially.

\section{Evaluation of synchronization error}

Now let us evaluate the total guaranteed synchronization error $Q=\sup\overline{\lim\limits_{t\to\infty}}\|e(t)\|$ where $\sup$ is taken over the set of transmission errors $\delta_y(t)$ not exceeding the level $\Delta$ in absolute value The ratio $C_e=Q/\Delta$ (the relative error) can be interpreted as the norm of the transformation from the input function $\delta_y(\cdot)$ to the output function $e(\cdot)$ generated by the system \eqref{10}. We will assume that the nonlinearity is Lipschitz continuous along all the trajectories starting from $\Omega$. More precisely, we assume that 
\begin{align*}
\|\ph(y)-\ph(y')\|\le L_\ph|y-y'|
\end{align*}
for $y=Cx$, $y'=Cx'$, $\dist(x,\Omega)\le \Delta$, $\dist(x',\Omega)\le \Delta$.

The error equation \eqref{10} can be represented as 
\begin{align}
\dot e=A_Le+\xi(t),
\label{11}
\end{align}
where $\|\xi(t)\|\le \big(L_\ph+\|L\|\big)\Delta$.
Choose $L$ such that $A_L$ is a Hurwitz (stable) matrix and choose a positive-definite matrix $P=P\trn>0$ satisfying the modified Lyapunov equation $PA_L+A_L\trn P\le -\mu P$, for some $\mu>0$. After simple algebra we obtain the differential inequality for the function $V(t)=e(t)\trn Pe(t)$:
\begin{align*}
\dot V\le -\mu V+e\trn P\xi(t)\le-\mu V+\sqrt{V}\cdot\sqrt{\xi\trn P\xi}.
\end{align*}

Since $\dot V\!<\!0$ within the set \mbox{$\sqrt{V}\!>\!\mu^{-1}\sup\limits_{t}\sqrt{\xi(t)\trn P\xi(t)}$,} the value of $\overline{\lim\limits_{t\to\infty}}\sup V(t)$ cannot exceed $\Delta^2\big(L_\ph+\|L\|\big)^2\lambda_{\max}(P)/\mu^2$. In view of positivity of  $P$, $\lambda_{\min}(P)\|e(t)\|^2\le V(t)$, where $\lambda_{\min}(P)$, $\lambda_{\max}(P)$ are minimum and maximum eigenvalues of $P$, respectively. Hence
\begin{align}
\overline{\lim\limits_{t\to\infty}}\|e(t)\|\le C_e^{+}\Delta,
\label{12}
\end{align}
where $C_e^{+}=\sqrt{\dfrac{\lambda_{\max}(P)}{\lambda_{\min}(P)}}\dfrac{L_\ph+\|L\|}{\mu}$. 

The relation \eqref{12} shows that the total synchronization error is proportional to the upper bound of transmission error $\Delta$, i.e. can be made arbitrarily small for sufficiently large transmission rate $R$.

One can pose the following problem: choose an optimal gain vector $L$ providing the minimum value of $C_e$. However an analytical solution is difficult to obtain in view of the system nonlinearity. An alternative approach is to evaluate upper and lower bounds for $C_e$ based on worst case inputs $\delta_y(t)$. Such a problem is similar to the energy control problem for systems with dissipation \cite{Fradkov05,Fradkov06} and $C_e$ can be interpreted as {\it excitability index} of the system. Employing the lower bound for excitability index for passive systems \cite{Fradkov05,Fradkov06} we conclude that if the gain vector $L$ is chosen to ensure strict passivity of the system \eqref{10} then the lower bound for $C_e$ is positive, i.e.
\begin{align}
\sup\limits_{|\delta_y(t)|\le\Delta}\overline{\lim\limits_{t\to\infty}}\|e(t)\|\ge C_e^{-}\Delta.
\label{12a}
\end{align}
Therefore for finite channel capacity the guaranteed synchronization error does not reduce to zero being of the same order of magnitude as the transmission error. 
Let us apply the above results to synchronization of two chaotic Chua systems coupled via a channel with limited capacity.

\section{Synchronization of chaotic Chua systems}

{\it System Equations}. Consider the chaotic Chua system model:
\begin{align}
\label{chuagen}
&\begin{cases}
\dot x_1=p(-x_1+\ph(y)+x_2),\quad t\ge 0,\cr
\dot x_2=x_1-x_2+x_3\cr
\dot x_3=-qx_2,
\end{cases}\\
&y(t)=x_1(t),\nonumber
\end{align}
where $y(t)$ is the sensor output (to  be transmitted over the communication channel), $p$, $q$ are 
known plant model parameters, $x=[x_1,x_2,x_3]\trn\inr^3$ is the plant state vector, the initial condition vector $x_0=x(0)$ is assumed to be unknown, $\ph(y)$ is a piecewise-linear function,
 having the following form:
\begin{align}
\ph(y)&= m_0y+|x+1|-|x-1|\cr
&+0.5(m_1-m_0)(|x+1|-|x-1|),
\label{phi}
\end{align}
where $m_0$, $m_1$ are given plant parameters.

{\it Observer design}. 
To obtain estimates $\hat x(t)$ of the current state $x(t)$ of the system \eqref{chuagen}, 
the special case of a continuous-time observer \eqref{10} is designed as follows
\begin{align}
\label{chuaobs}
&\begin{cases}
\dot {\hat{x}}_1=p(-\hat{x}_1+\ph(y)+ \hat{x}_2)+l_1\eps(t),\cr
\dot {\hat{x}}_2=\hat{x}_1-\hat{x}_2+\hat{x}_3+l_2\eps(t),\cr
\dot {\hat{x}}_3=-q\hat{x}_2+l_3\eps(t),\cr
\eps(t) =\bar y(t)- \hat{y}(t),
\end{cases}\\
&\hat{y}(t)=\hat{x}_1(t),\quad \hat{x}(0)= \hat{x}_0, \nonumber
\end{align} 
 where   $l_1$, $l_2$, $l_3$ are observer parameters, forming the $3\times 1$ observer matrix gain $L=[l_1,l_2,l_3]\trn$.
 
Subtracting  \eqref{chuaobs} from \eqref{chuagen} yields 
\begin{align}
\label{obserr}
&\begin{cases}
\dot {e}_1=p(-e_1+ e_2)+l_1\big(\delta_y(t)-e_1\big)+\xi_1(t),\cr
\dot {e}_2=e_1-e_2+e_3+ l_2\big(\delta_y(t)-e_2\big),\cr
\dot {e}_3=-qe_2+ l_3\big(\delta_y(t)-e_3\big),\cr
\xi_1(t)=\ph\big(y(t)-\delta_y(t)\big)- \ph\big(y(t)\big).
\end{cases}
\end{align} 

Equation \eqref{obserr} describes the linear time-invariant (LTI) system 
$\dot e(t)=Ae(t)$, $e(0)=x_0-\hat{x}_0$ with the following matrix $A$:
\begin{align}
\label{matra}
A=\begin{bmatrix}
-p-l_1&p&0\cr
1-l_2&-1&1\cr
-l_3&q&0
\end{bmatrix}.
\end{align} 

Matrix $L$ should be chosen so that the observer \eqref{chuaobs} stability conditions are satisfied, i.e. the characteristic polynomial $D_L(s)=\det(s{\bf I}-A_L)$ is Hurwitz. 
For the observer \eqref{chuaobs}, the polynomial $D_L(s)$ has the form:
\begin{align}
\label{dots}
D_L(s)&=s^3+(1+p+l_1)s^2+(-q+pl_2+l_1)s\nonumber\\
&-pq+l_3p-l_1q.
\end{align}

Evidently, we may find the matrix $L$ for any arbitrarily assigned parameters $d_1$, $d_2$, $d_3$ 
so that the characteristic polynomial $D_L(s)=s^3+d_1s^2+d_2s+d_3$. This leads to asymptotic convergence of the synchronization error $e(t)$ to zero with  prescribed dynamics in the disturbance-free case.

\begin{figure}[htpb]
\centering
\includegraphics[width=75mm]{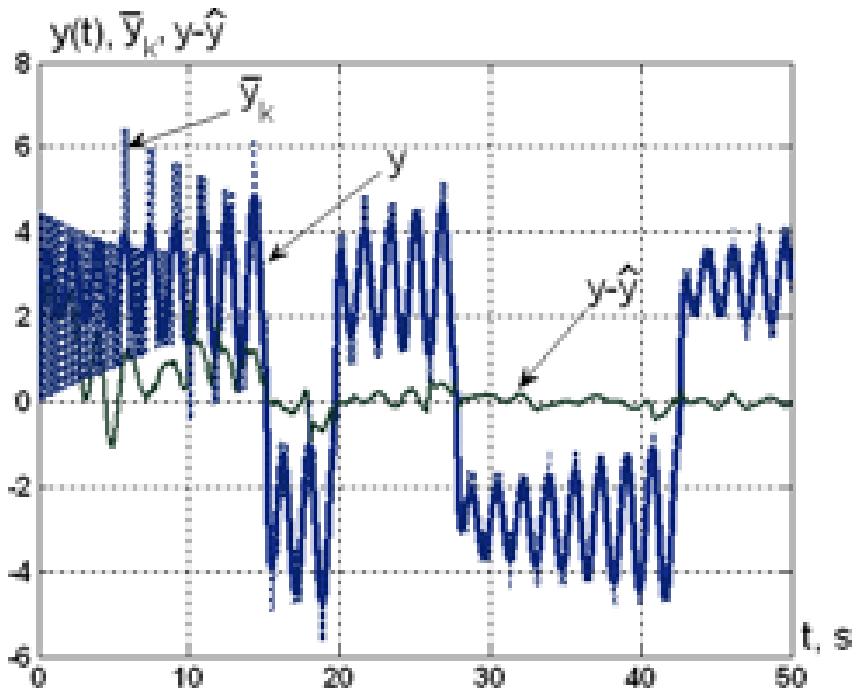}
\caption{Outputs $y(t)$, $\hat{y}(t)$, $y\bar (t_k)$ time histories; $\Delta=1$.}
\label{ydy}
%\end{figure}
%\begin{figure}[htpb]
\centering
\includegraphics[width=75mm]{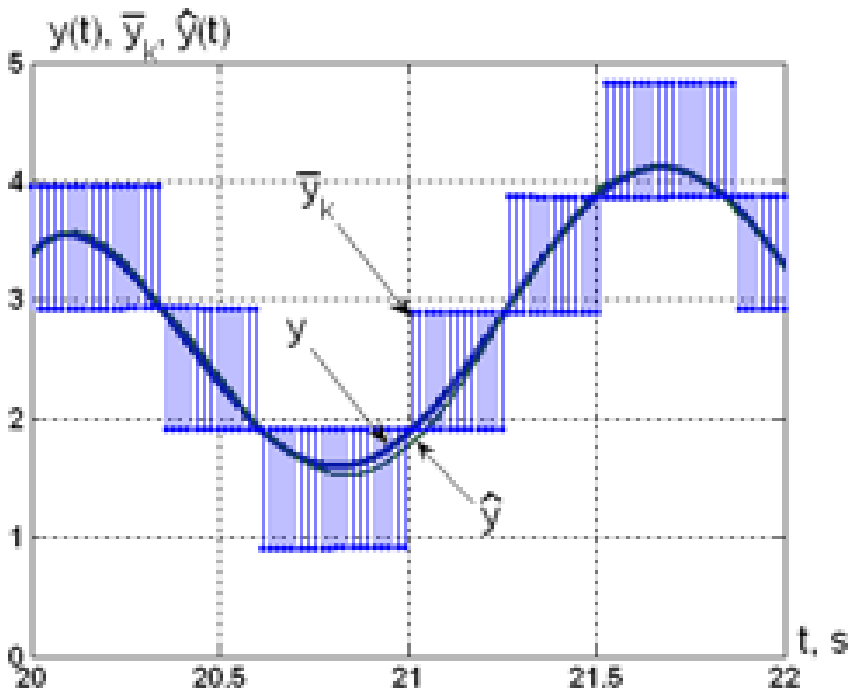}
\caption{Zooming of Fig.\protect\ref{ydy} for $t\in[20,22]$~s.}
\label{ydyzoom}
%\end{figure}
%\begin{figure}[htpb]
\centering
\includegraphics[width=75mm]{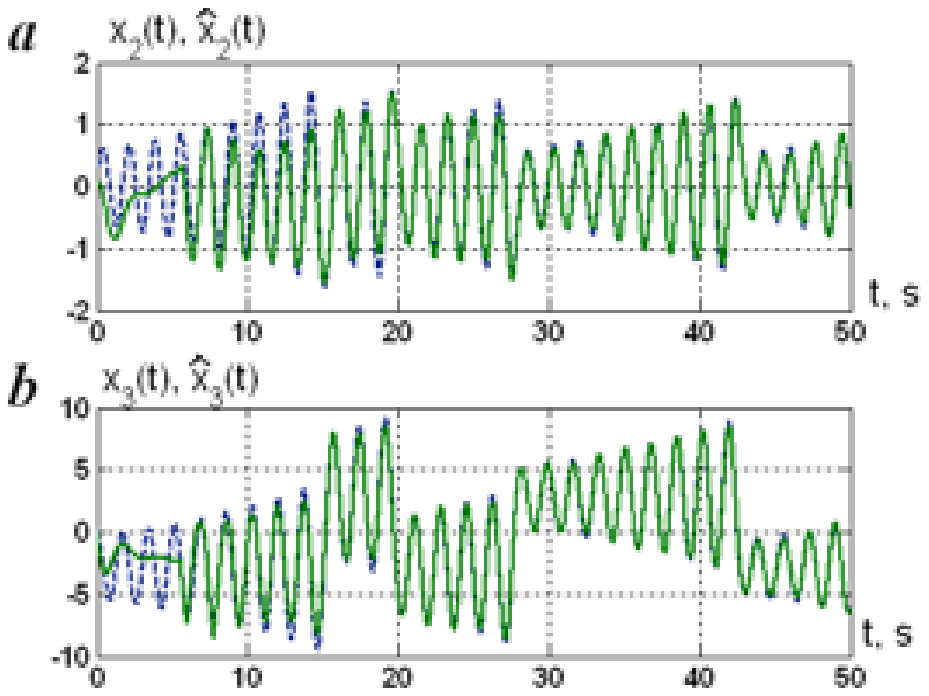}
\caption{Time histories for $\Delta=1$: {\it a}) $x_2(t)$ (- - -), $\hat{x}_2(t)$ (---); 
{\it b}) $x_3(t)$ (- - -), $\hat{x}_3(t)$ (---). }
\label{xdx}
\end{figure}

\begin{figure}[htpb]
\centering
\includegraphics[width=75mm]{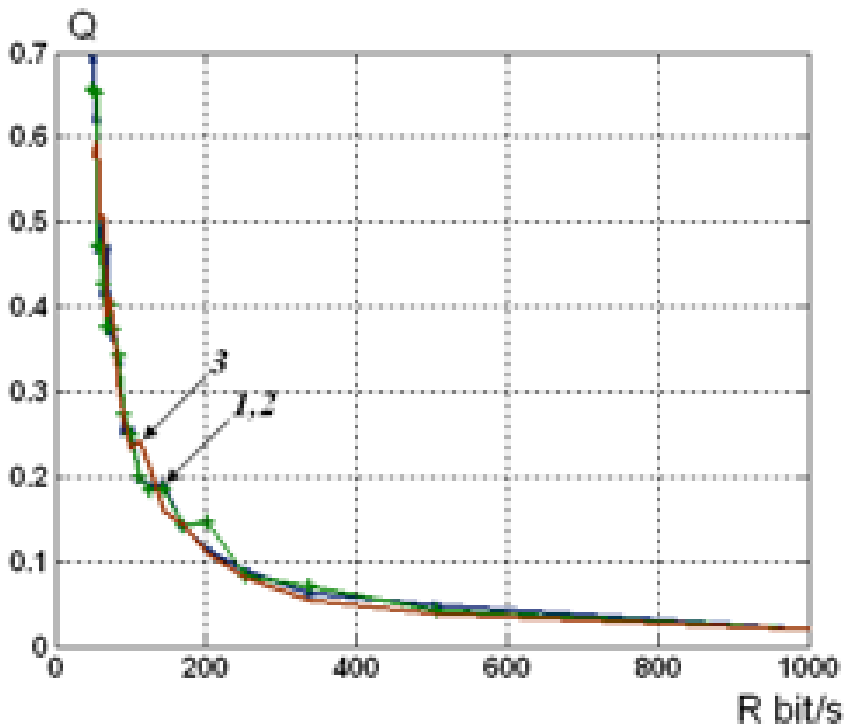}
\caption{Synchronization error $Q$ vs $\Delta$ for different $L$.}
\label{qotdelta1}
%\end{figure}
%\begin{figure}[htpb]
\centering
\includegraphics[width=75mm]{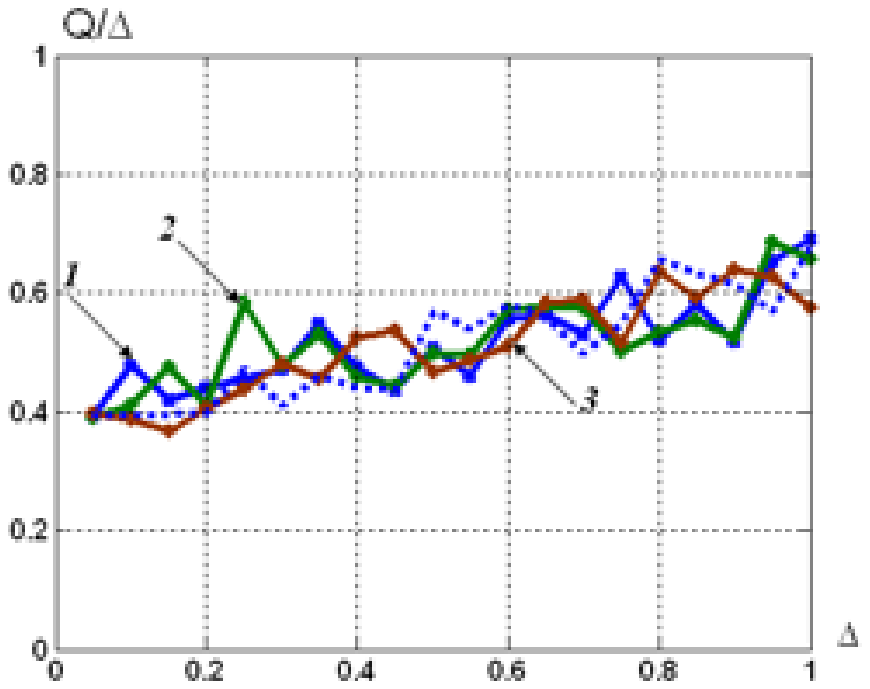}
\caption{Relative synchronization error $Q/\Delta$ vs $\Delta$ for different $L$.}
\label{qotdivdelta1}
%\end{figure}
%\begin{figure}[htpb]
\centering
\includegraphics[width=75mm]{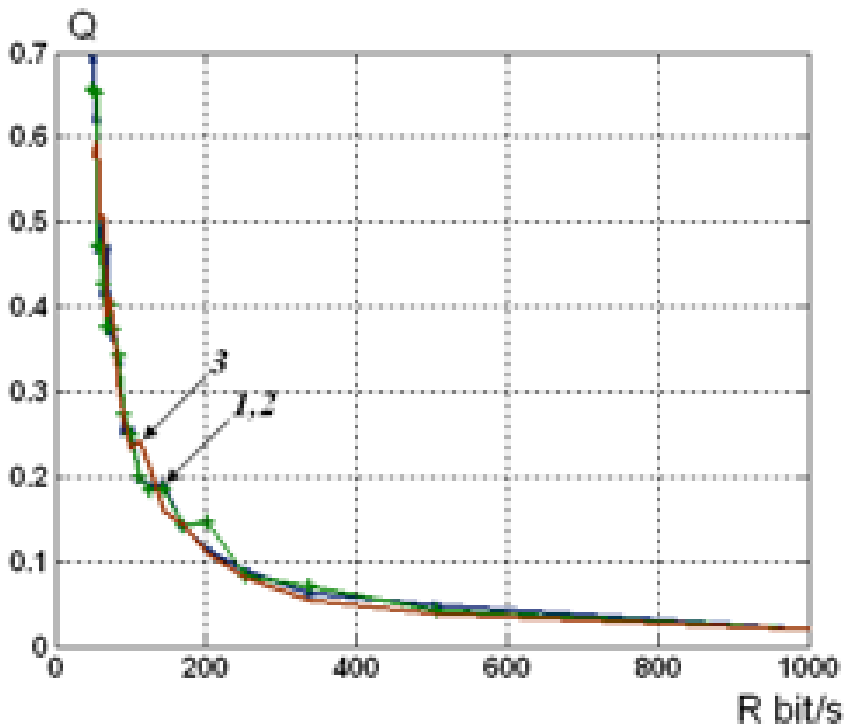}
\caption{Synchronization error $Q$ vs transmission rate $\bar R$.}
\label{qotr1}
\end{figure}

%\subsection{Simulation results}
{\it Simulation results}. For simulation the following parameter values of the Chua system model \eqref{chuagen} were chosen: $p=10.0$, $q=15.6$,
$m_0=0.33$, $m_1=0.22$. The system exhibits a chaotic behavior, see $y(t)$ in  Fig.~\ref{ydy}.

In our simulations parameter $\Delta$ has been taken from the set $\Delta=\{0.1, 0.2, 0.5, 0.7, 1\}$. The sampling time $T_s$ for each $\Delta$ has been chosen in accordance with \eqref{topt} for $L_y=30$~s$^{-1}$. In \eqref{mk} the initial value $M_0$ has been taken as $ M_0=5$, decay parameter $\rho=\exp(-0.1T_s)$, and limit value $M_\infty=M^*=\Delta/2$.

To evaluate the minimal synchronization error the optimal observer gain matrix $L^*(\Delta)$ was found numerically for several values of the transmission error $\Delta$. We obtained $L^*(0.1)\!= \![-4.66, 0.50, -4.40]\trn $, $L^*(0.5)\!=\! [\!-\!4.40, 0.46, \!-\!4.54]\trn $ $L^*(1.0)\!= \![\!-\!4.97, 0.46, \!-\!4.47]\trn $. For comparison the observer design by assigning a {\it Butterworth} distribution of the observer  matrix $A_L$ eigenvalues was performed. For the third order system the Butterworth design provides characteristic polynomial \eqref{dots} as $D(s)=s^3+2\omega_0s^2+
2\omega_0^2s+\omega_0^3$, where parameter $\omega_0>0$ specifies the desired estimation rate. In our example $\omega_0=6$~s$^{-1}$ is taken. It provides the observer eigenvalues: $s_1=-6.0$, $s_{2,3}=-3.0\pm 5.2~i$.  
The observer feedback gain matrix $L$ is found as $L=[1.00, 5.54, 4.44]\trn$.  
For simulation the initial condition vectors for the systems \eqref{chuagen} and \eqref{chuaobs}
were taken as $x_0=[0.3,0.3,0.3]\trn$ and $\hat{x}_0=[0, 0, 0]\trn$.

Simulation results for the coder \eqref{qu1}, \eqref{dybar}, \eqref{mk} and observer with optimally chosen gains for $\Delta=1$ are shown in Figs.~\ref{ydy},~\ref{ydyzoom}, \ref{xdx}.  The sampling interval is $T_s=0.02$~s, which corresponds to the transmission rate $\bar R=50$ bits per second. The following coder parameters were  chosen: $M_0=5.0$, $M_\infty=0.5$, $\rho= 0.998$. It is seen that the synchronization process possesses sufficiently fast dynamics even in the presence of information constraints.

It is seen from Fig.~\ref{qotdelta1} that the difference in the limit synchronization error for different rationally chosen observer gains is not significant. Moreover, it is seen from Fig.~\ref{qotdivdelta1} that the relative error does not approach zero for all choices of the observer gains.

Dependence of the synchronization error $Q$ on the transmission rate $\bar R$ is shown in Fig.~\ref{qotr1}, demonstrating that the synchronization error becomes small for sufficiently large transmission rates.

\section{Conclusions}
We have studied dependence  of the synchronization error in the observer-based synchronization system both analytically and numerically. It is shown that upper and lower bounds for limit synchronization error depend linearly on the transmission error which, in turn, is proportional to the driving signal rate and inversely proportional to the transmission rate. Though these results are obtained for a special type of coder, it reflects peculiarity of the synchronization problem as a nonequilibrium dynamical problem. On the contrary, the stabilization problem considered previously in the literature on control under information constraints belongs to a class of equilibrium problems.
As an intermediate result we obtained relation \eqref{ropt2} playing the role of an uncertainty relation between the transmission rate of information and the transmission error.

Future research is aimed at analysis of controlled synchronization and control of chaos problems under information constraints. 

\medskip

The work was supported by NICTA, University of Melbourne and Russian Foundation for Basic Research (project RFBR 05-01-00869).

\bibliography{FAE_PhRL05}
\end{document}